\documentclass[a4paper,11pt,fleqn]{article}

\usepackage{amsmath,amssymb,amsthm,cite} 
\usepackage{rotating}
\usepackage{lscape}

\newcommand{\todo}[1][\null]{\ensuremath{\clubsuit}}
\newcommand{\checked}[1][\null]{\ensuremath{\diamond}}
\newcommand{\noprint}[1]{}

\newcounter{tbn}

\newcounter{mcasenum}

\newtheorem{theorem}{Theorem}

\newtheorem*{proposition*}{Proposition}
{\theoremstyle{definition}

\newtheorem{note}{Note}
}

\begin{document}
\begin{center}
\par\noindent {\LARGE\bf Enhanced group classification  of \\Benjamin--Bona--Mahony--Burgers equations

\par}

{\vspace{5mm}\par\noindent\large Olena~Vaneeva$^{\dag 1}$, Severin Po\v sta$^{\ddag 2}$ and Christodoulos Sophocleous$^{\S 3}$
\par\vspace{2mm}\par}
\end{center}

{\par\noindent\it\small
${}^\dag$\ Institute of Mathematics of NAS of Ukraine,
 3 Tereshchenkivska Str., 01601 Kyiv-4, Ukraine \\[1ex]
${}^\ddag$\ Department of Mathematics, Faculty of Nuclear Sciences and Physical Engineering,\\[1ex]
$\phantom{{}^\ddag}$\ Czech Technical University in Prague, 13 Trojanova Str., 120 00 Prague, Czech Republic\\[1ex]
${}^\S$\ Department of Mathematics and Statistics, University of Cyprus, Nicosia CY 1678, Cyprus
}

{\vspace{3mm}\par\noindent
$\phantom{{}^\dag{}\;}\ $E-mails: \it $^1$vaneeva@imath.kiev.ua, $^2$severin.posta@fjfi.cvut.cz, $^3$christod@ucy.ac.cy

\par}

{\vspace{5mm}\par\noindent\hspace*{5mm}\parbox{150mm}{\small
A class of the Benjamin--Bona--Mahony--Burgers (BBMB) equations with time-dependent coefficients is investigated with the Lie symmetry point of view.
The set of admissible transformations of the class is described exhaustively.
The complete group classification is performed using  the method of mapping between classes.
The derived Lie symmetries are used to reduce BBMB equations to ordinary differential equations.
 Some exact solutions are constructed.
}\par\vspace{4mm}}

\section{Introduction}

The regularized long wave equation $u_t+u_x+uu_x-u_{xxt}=0$ was proposed by Peregrine~\cite{peregrine} and later by Benjamin et
al.~\cite{Benjamin}
to describe small-amplitude long
waves on the surface of water in a channel.  In order to take into account the mechanisms leading to the degradation of the wave
the model including dissipative term $u_{xx}$ was considered in~\cite{Amick1989}, namely
\begin{equation*}\label{bbmb}
u_t+u_x+uu_x-\nu u_{xx}-u_{xxt}=0, \quad \nu\in\mathbb{R}^+.
\end{equation*}  Here $u = u(x, t)$ is a real-valued function
of the two real variables $x$ and $t$, which, in applications, are typically
proportional to distance in the direction of propagation and to elapsed
time, respectively. The dependent variable may represent a displacement
of the underlying medium or a velocity~\cite{Amick1989}.  The regularized long wave equation with a Burgers-type dissipative term appended  is more frequently  called  the Benjamin--Bona--Mahony--Burgers (BBMB) equation~\cite{Naumkin}.

Last decade model equations with time-dependent coefficients are intensively studied with different points of view. This is due to the fact that they can model real phenomena with more accuracy taking into account parameters changing in time, like, for example, slightly varying depth of water or thickness of ice layer~\cite{kuriksha}. The most general form of the BBMB equation with time-dependent coefficients is
\begin{equation}\label{bbmb_fghk}
u_t+f(t)u_x+g(t)uu_x+k(t)u_{xx}+h(t)u_{xxt}=0,\quad ghk\neq0,
\end{equation}
where $f$, $g$, $h$, and $k$ are smooth functions of the variable $t$.

In this paper we aim to investigate this class with the Lie symmetry point of view, namely, to
 present the complete group classification of this class of equations.
A similar study was initiated in~\cite{Kumar} but the complete and correct group classification was not achieved therein.
Lie symmetries and conservation laws of equations~\eqref{bbmb_fghk} without dissipative term (i.e. with $k=0$) were thoroughly investigated in~\cite{VPS2015}.

There are several reasons why it is important to have the complete and correct group classification of a class of nonlinear PDEs.
Firstly, the classification list reveals all  equations from the class under study which can be reduced to PDEs with fewer number of independent variables or even to ODEs. Then the powerful and, what is most important, algorithmic Lie reduction method can be used to construct exact solutions~\cite{olve1993b,Ovsiannikov,Ibragimov}. Secondly, the equations admitting Lie symmetry extensions potentially are more interesting for applications~\cite{FN}. There are also other advantages of  Lie symmetries such as generation of new solutions from known ones, construction of conservation laws, etc. It should be emphasized that though there exist several computer algebra packages for computation of Lie symmetries, and finding of maximal Lie symmetry algebra of a certain fixed differential equation is a routine task which can be done automatically in majority of cases this is not the case for the group classification problem. Though there exist successful examples for some simple classes involving a few number of arbitrary elements, for the rest of cases the involvement of a researcher and recently developed tools of group analysis are needed.

We carry out the group classification of class~\eqref{bbmb_fghk} using the method of mapping between classes suggested in~\cite{vane2009} and then successfully applied for several classes of variable coefficient PDEs, see, e.g., recent works~\cite{VKS2015,VPS2015}. When the complete group classification is achieved Lie reductions of BBMB equations to ODEs are performed as well as some exact solutions are constructed. It is understood that not all group-invariant solutions will fit the specific physical situation and respective boundary conditions, this problem, in particular, was discussed recently in~\cite{pucci}. Nevertheless, even ``non-physical'' exact solutions can be usefull, for example, for testing numerical methods.

\section{Group classification}
The statement of the group classification problem is formulated as follows: given a class of differential equations, the goal is to classify all possible cases of extension of Lie invariance
algebras of such equations with respect to the equivalence group of the class~\cite{Ovsiannikov}.

Consider firstly the transformational properties of class~\eqref{bbmb_fghk}. We look for the admissible  point transformations (triples consisting of two equations from the class and a nondegenerate point transformations linking them~\cite{popo2010a}) using the direct method~\cite{kingston1997}.
 The proofs of the statements below are similar to those presented in~\cite{VPS2015} for equations~\eqref{bbmb_fghk} with $k=0$. Thus, we skip the details of calculations for the sake of brevity and present the final results only.

All the admissible transformations in class~\eqref{bbmb_fghk} are generated by equivalence transformations and therefore this class is normalized in the usual sense.  The following  statement is true.
\begin{theorem}
The usual equivalence group~$G^{\sim}$ of class~\eqref{bbmb_fghk} consists of the transformations
\begin{gather*}
\tilde t=T(t),\quad \tilde x=\delta_1x+\delta_2,  \quad
\tilde u=\delta_3u+\delta_4, \\
\tilde k(\tilde t)=\dfrac{\delta_1{}^2}{T_t}k(t),\quad
\tilde f(\tilde t)=\dfrac{\delta_1}{T_t\delta_3}(\delta_3 f(t)-\delta_4g(t)), \quad
\tilde g(\tilde t)=\dfrac{\delta_1}{T_t\delta_3}g(t),\quad
\tilde h(\tilde t)=\delta_1{}^2h(t),
\end{gather*}
where $\delta_j$, $j=1,2,3,4$, are arbitrary constants with $\delta_1\delta_3\not=0$
and $T=T(t)$ is an arbitrary smooth function with $T_t\neq0.$
Class~\eqref{bbmb_fghk} is normalized in the usual sense.
\end{theorem}

Using this theorem, we can formulate the criterion of reducibility of variable coefficient BBMB equations from class~\eqref{bbmb_fghk} to their constant coefficient counterparts:

{\it
A variable-coefficient equation from class~\eqref{bbmb_fghk} is reduced to a constant-coefficient equation from the same class  by a point transformation
if and only if
the corresponding coefficients $f,$ $g$, $h$, and $k$ satisfy the conditions
\begin{equation*}\label{criterion1}
\left( f/g\right)_t=h_t=k_t=0.
\end{equation*}
i.e., $k$ and $h$ are constants and the function $f$ is proportional to $g$.}

We note that the maximal Lie invariance algebra $A^{\rm max}$ of the constant coefficient BBMB equation was found in~\cite{Bruzon2009}. It is two-dimensional Abelian algebra $\langle\partial_t,\partial_x\rangle$ spanned by the operators of time and space translations.

\begin{table}[ht]
\caption{\label{TableLieSym1}The group classification of  class~\eqref{eqBBMimaged} up to $G^\sim_1$-equivalence.}
\begin{center}
\begin{tabular}{ccccl}
\hline
no.&$H(t)$&$K(t)$&$F(t)$&\hfil Basis of $A^{\max}$ \\
\hline
0&$\forall$&$\forall$&$\forall$&$\partial_x$\\
1&$\varepsilon t^\rho$&$\lambda t^{\rho-1}$&$\delta t^{\frac{\rho-4}{2}}$&
$\partial_x,\ 2t\partial_t+\rho x\partial_x+(\rho-2)u\partial_u$\\
2&$\varepsilon e^t$&$\lambda e^t$&$\delta e^{\frac12t}$&
$\partial_x,\  2\partial_t+x\partial_x+u\partial_u$\\
3&$\varepsilon$&$\lambda$&$\delta$&
$\partial_x,\ \partial_t$\\
\hline
\end{tabular}
\\[1ex]
\parbox{120mm}{Here $\varepsilon$, $\delta$, $\lambda$, and $\rho$ are arbitrary constants with $\varepsilon\lambda\neq0$, $\varepsilon=\pm1\bmod G^\sim_1$. In Case 3 $\delta=0,1\bmod G^\sim_1$ and additionally $\lambda=-1\bmod G^\sim_1$ if $\delta=0$. }
\end{center}\end{table}

The presence of four arbitrary elements in class~\eqref{bbmb_fghk} leads to difficulties in solving the group classification problem. Therefore,
we firstly simplify the problem by reducing the number of arbitrary elements in the class. This can be done either via gauging of arbitrary elements by equivalence transformations or using the method of mapping between classes (see~\cite{VPS2015} for the details).  We choose the second option.

The family of point transformations
\begin{equation}\label{trr}
\tilde t=\int\!\! g(t){\rm d}t,\quad \tilde x=x,\quad\tilde u=u+\frac{{f}(t)}{g(t)},
\end{equation}
parameterized by two arbitrary elements of class~\eqref{bbmb_fghk}, maps
class~\eqref{bbmb_fghk}  to the related class of variable coefficient BBMB equations with a forcing term
\begin{equation}\label{eqBBMimaged}
u_t+ uu_x+K(t)u_{xx}+ H(t)u_{xxt}=F(t),\quad HK\neq0.
\end{equation}
(Tildes in~\eqref{eqBBMimaged} are omitted.)
The arbitrary elements of the initial class~\eqref{bbmb_fghk} and  the imaged class~\eqref{eqBBMimaged} are related via the formulas
\begin{equation}\label{form_arb_el}
K(\tilde t)=\frac{k(t)}{g(t)},\quad H(\tilde t)=h(t),\quad F(\tilde t)=\frac1{g(t)}\left(\frac{f(t)}{g(t)}\right)_t.
\end{equation}
Following the method of mapping between classes, we firstly classify Lie symmetries of the imaged class~\eqref{eqBBMimaged}
and then use the family of point transformations~\eqref{trr} and the relations \eqref{form_arb_el} to extend the result to the initial class~\eqref{bbmb_fghk}.

In order to efficiently solve the group classification problem for class~\eqref{eqBBMimaged},
we look for admissible transformations in this class using the direct method.
Similarly to class~\eqref{bbmb_fghk} such transformations are exhausted by transformations from the usual equivalence group admitted by this class.
\begin{theorem}
The usual equivalence group~$G^{\sim}_1$ of class~\eqref{eqBBMimaged} comprises the transformations
\begin{gather*}
\tilde t=\dfrac{\delta_1}{\delta_3}t+\delta_0,\quad \tilde x=\delta_1x+\delta_2,  \quad
\tilde u=\delta_3u,\\
\tilde K(\tilde t)=\delta_1\delta_3 K(t),\quad
\tilde H(\tilde t)=\delta_1{}^2H(t),\quad
\tilde F(\tilde t)=\dfrac{\delta_3{}^2}{\delta_1}F(t),
\end{gather*}
where $\delta_j$, $j=0,1,2,3$, are arbitrary constants with $\delta_1\delta_3\not=0$.
Class~\eqref{eqBBMimaged} is normalized in the usual sense.
\end{theorem}
Further we use Lie infinitesimal invariance criterion and get the determining equations involving coefficients of symmetry generators $\Gamma$, arbitrary elements of the class and derivatives of both entities. Integration of those determining equations that do not involve arbitrary elements leads

\begin{landscape}
\begin{table}[h!]
\begin{center}
\caption{\label{TableLieSym2}The group classification of class~\eqref{bbmb_fghk}
up to $G^\sim$-equivalence.}
\begin{tabular}{ccccl}
\hline
no.&$h(t)$&$k(t)$&$f(t)$&\hfil Basis of $A^{\max}$ \\
\hline
0&$\forall$&$\forall$&$\forall$&$\partial_x$\\
1&$\varepsilon t^\rho$&$ \lambda t^{\rho-1}$&$\delta t^{\frac{\rho-2}{2}}$&
$\partial_x,\ 2t\partial_t+\rho x\partial_x+(\rho-2)u\partial_u$\\
2&$\varepsilon t^2$&$\lambda t$&$\delta\ln t$&
$\partial_x,\ t\partial_t+x\partial_x-\delta\partial_u$\\
3&$\varepsilon e^t$&$\lambda e^t$&$\delta e^{\frac12t}$&
$\partial_x,\  2\partial_t+x\partial_x+u\partial_u$\\
4&$\varepsilon$&$\lambda$&$\delta t$&
$\partial_x,\ \partial_t-\delta\partial_u$\\
\hline
\end{tabular}
\\[1ex]
\parbox{150mm}{Here $g(t)=1\bmod G^\sim$; $\varepsilon$, $\delta$, $\lambda$, and $\rho$ are arbitrary constants with $\varepsilon\lambda\neq0$, $\varepsilon=\pm1\bmod G^\sim$. In Case 4 $\delta=0,1\bmod G^\sim$ and additionally $\lambda=-1\bmod G^\sim$ if $\delta=0$.}
\end{center}
\end{table}
\begin{table}[h!]
\begin{center}
\caption{\label{TableLieSym3}The group classification of class~\eqref{bbmb_fghk}
using no equivalence.}
\begin{tabular}{ccccl}
\hline
no.&$h(t)$&$k(t)$&$f(t)$&\hfil Basis of $A^{\max}$ \\
\hline
0&$\forall$&$\forall$&$\forall$&$\partial_x$
\\\rule{0ex}{4.2ex}
1&$\mu_1(\varepsilon T+\kappa)^\rho$&$\mu_2g(\varepsilon T+\kappa)^{\rho-1}$& $\mu_3g(\varepsilon T+\kappa)^{\frac{\rho-2}2}+\mu_4g$&
$\partial_x,\ \dfrac2g(\varepsilon T+\kappa)\partial_t+\varepsilon\rho x\partial_x+\varepsilon(\rho-2)(u+\mu_4)\partial_u$
\\\rule{0ex}{4.2ex}
2&$\mu_1(\varepsilon T+\kappa)^2$&$\mu_2g(\varepsilon T+\kappa)$&$\mu_3g\ln(\varepsilon T+\kappa)+\mu_4g$&
$\partial_x,\  \dfrac1g(\varepsilon T+\kappa)\partial_t+\varepsilon x\partial_x-\varepsilon\mu_3\partial_u$
\\\rule{0ex}{4.2ex}
3&$\mu_1 \exp({\sigma T})$&$\mu_2g \exp({\sigma T})$&$\mu_3 g\exp({\frac12\sigma T})+\mu_4g$&
$\partial_x,\  \dfrac2g\partial_t+\sigma x\partial_x+\sigma(u+\mu_4)\partial_u$
\\\rule{0ex}{4.2ex}
4&$\mu_1$&$\mu_2g$&$\mu_3gT+\mu_4g$&$\partial_x,\ \dfrac1g\partial_t-\mu_3\partial_u$
\\
\hline
\end{tabular}
\\[1ex]
\parbox{170mm}{Here $g$ is an arbitrary nonvanishing smooth function, $T=\int\!g(t)\,{\rm d}t$; $\varepsilon=\pm1$;
$\mu_i$, $i=1,\dots,4,$ $\sigma$, $\kappa$ and~$\rho$ are arbitrary constants with $\sigma\mu_1\mu_2\neq0$.}
\end{center}
\end{table}

{\small
\begin{note}
Any constant coefficient BBMB equation $u_t+fu_x+guu_x+k u_{xx}+h u_{txx}=0$ can be  reduced to the equation $u_t+uu_x-u_{xx}+\varepsilon u_{txx}=0$,\\ where $\varepsilon=\mathrm{sign}(h)$, (Case 4 of Table~2 with $\delta=0$ and $\lambda=-1$) by the transformation $\tilde t=-\dfrac{k}{{|h|}}t,$ $\tilde x=\dfrac{x}{\sqrt{|h|}},$ $\tilde u=-\dfrac{\sqrt{|h|}}{k}(gu+f)$ from the group~$G^\sim$.
\end{note}}
\end{landscape}

\newpage \noindent to the general form of the infinitesimal generators  $\Gamma=(c_1t+c_0)\partial_t+(c_2x+c_3)\partial_x+(c_2-c_1)u\partial_u,$ where $c_i,$ $i=0,\dots3,$ are arbitrary constants. Then the  classifying equations involving arbitrary elements take the form
\[(c_1t+c_0)H_t=2c_2H,\quad (c_1t+c_0)K_t=(2c_2-c_1)K,\quad (c_1t+c_0)F_t=(c_2-2c_1)F.\]
We integrate these equations up to the $G^\sim_1$-equivalence and obtain the classification list. The results are summarized in Table~\ref{TableLieSym1}.

The classification list for the initial class~\eqref{bbmb_fghk} can be obtained using the transformation~\eqref{trr}, the relations \eqref{form_arb_el}
and the group classification results derived for the imaged class~\eqref{eqBBMimaged} (Table~1). The transformation~\eqref{trr}
can be considered as a composition of the equivalence transformation $\mathcal\tau^\sim\colon\ \tilde t=\int\!\! g(\bar t){\rm d}\bar t$, $\tilde x=\bar x$, $\tilde u=\bar u$
from the group $G^\sim$ and the transformation $\tau\colon \ \bar t=t$, $\bar x=x$, $\bar u=u+\frac{{f}(t)}{g(t)}$ that does not belong to $G^\sim$.
The transformation $\mathcal\tau^\sim$ maps any equation from (1) to the equation from the same class with $g=1.$ In order to obtain the group classification
for class~\eqref{bbmb_fghk} up to the $G^\sim$-equivalence it is enough to consider the transformation~\eqref{trr} with $g=1.$
Then the formulas~\eqref{form_arb_el} which connect  arbitrary elements in classes~\eqref{bbmb_fghk} and~\eqref{eqBBMimaged} take the simple form $H=h$, $K=k$, $F=f_t.$ We integrate the latter ODE for the values of $F$ appearing in Table~1. We, respectively, get the following forms of $f=f(t)$:

1. $f=\bar\delta t^{\frac{\rho-2}{2}}+C$ ($\bar\delta=2\delta/(\rho-2)$), if $\rho\neq2$ and $f=\delta \ln t+C,$ otherwise;

2. $f=\bar\delta e^{\frac12t}+C$ ($\bar \delta=2\delta$);

3. $f=\delta t+C$.

\noindent
The integration constant $C=0\bmod G^\sim.$
The last step is to perform the change of variable $\tilde u=u+f(t)$ in basis operators of the maximal Lie invariance algebras presented in Table~1. The results are summarized in Table~2.

We also derive the complete list of Lie symmetry extensions for the entire class~\eqref{bbmb_fghk},
where arbitrary elements are not simplified by point transformations (Table~3).
The group classification results reveal equations of the form~\eqref{bbmb_fghk} that may be of more interest for applications and  for which the classical  Lie reduction method can be  used.

\section{Reductions and exact solutions}
The classes of BBMB equations~\eqref{bbmb_fghk} and~\eqref{eqBBMimaged} are similar with respect to the transformations~\eqref{trr}.
If one have exact solutions for equations~\eqref{eqBBMimaged} the similar solutions for equations~\eqref{bbmb_fghk} can easily be recovered using~\eqref{trr}.
That is why it is convenient to perform classification of Lie reductions for class~\eqref{eqBBMimaged}, where the number of inequivalent cases of Lie symmetry extension is smaller.

To perform the classification of Lie reductions we need the optimal systems of one-dimensional subalgebras of the maximal
Lie invariance algebras of the BBMB equations~\eqref{eqBBMimaged}. Such algebras are at most two-dimensional. The optimal system of a two-dimensional Lie algebra
$\langle X_1, X_2\rangle$ is $\{\langle X_1\rangle, \langle X_2\rangle\}$ if the algebra is non-Abelian and $\{\langle X_1\rangle, \langle X_2+\alpha X_1\rangle\}$, where $\alpha\in\mathbb R$, if it is Abelian. We have non-Abelian algebras in Case~1 with $\rho\neq0$ and Case~2 of Table~1 and Abelian algebras in
Case~1 with $\rho=0$ and Case~3 of this table. We note that reductions with respect to subalgebra $\langle X_1=\partial_x\rangle$ lead to trivial constant solutions. Therefore, we perform only the reductions with the second subalgebra.
For each case we list the  BBMB equation, the one-dimensional subalgebra, the Ansatz constructed with this subalgebra and the corresponding reduced ODE.

Case $1_{\rho\neq0}$.
\begin{gather}\label{eqBBM1}
u_t+ uu_x+\lambda t^{\rho-1} u_{xx}+\varepsilon t^\rho u_{xxt}=\delta t^{\frac{\rho-4}{2}},\\\nonumber
\langle 2t\partial_t+\rho x\partial_x+(\rho-2)u\partial_u\rangle,\quad u=t^{\frac\rho2-1}\varphi(\omega),\ \mbox{where}\ \omega=xt^{-\frac\rho2},\\\nonumber
\varepsilon\rho\,\omega\varphi'''+(\varepsilon(\rho+2)-2\lambda)\varphi''-2\varphi\varphi'+\rho\,\omega\varphi'+(2-\rho)\varphi+2\delta=0.
\end{gather}

This ODE has particular exact solution $\varphi=\omega+\frac{2\delta}\rho,$ which gives the ``degenerate'' solution
$u=\dfrac xt+\dfrac{2\delta}\rho t^{\frac\rho2-1}$
of equation~\eqref{eqBBM1} with $\rho\neq0.$

Case $1_{\rho=0}$.
\begin{gather}\label{eqBBM1a}
u_t+ uu_x+\lambda t^{-1} u_{xx}+\varepsilon u_{xxt}=\delta t^{-2},\\\nonumber
\langle t\partial_t+\alpha\partial_x-u\partial_u\rangle,\quad u=\frac1t\varphi(\omega),\ \mbox{where}\ \omega=x-\alpha\ln t,\\\nonumber
\varepsilon\alpha \varphi'''+(\varepsilon-\lambda)\varphi''-\varphi\varphi'+\alpha\varphi'+\varphi+\delta=0.
\end{gather}
For $\alpha=-\delta$ this ODE has particular solution $\varphi=\omega+c$, where $c$ is an arbitrary constant.
This leads to the ``degenerate'' solution $u=\dfrac1t(x+\delta\ln t+c)$
of the equation~\eqref{eqBBM1a}.

Case~2.
\begin{gather}\label{eqBBM3}
u_t+ uu_x+\lambda e^t u_{xx}+\varepsilon e^t  u_{xxt}=\delta e^{\frac12t},\\\nonumber
\langle 2\partial_t+x\partial_x+u\partial_u\rangle,\quad u=e^{\frac t2}\varphi(\omega),\ \mbox{where}\ \omega=xe^{-\frac t2},\\\nonumber
\varepsilon\omega\varphi'''+(\varepsilon-2\lambda)\varphi''-2\varphi\varphi'+\omega\varphi'-\varphi+2\delta=0.
\end{gather}

Case~3.
\begin{gather}\label{eqBBM4}
u_t+ uu_x+\lambda  u_{xx}+\varepsilon  u_{xxt}=\delta,\\\nonumber
\langle \partial_t+\alpha\partial_x\rangle,\quad u=\varphi(\omega),\ \mbox{where}\ \omega=x-\alpha t,\\\nonumber
\varepsilon\alpha\varphi'''-\lambda\varphi''-\varphi\varphi'+\alpha\varphi'+\delta=0.
\end{gather}
Up to the equivalence we can consider  $\lambda=-1$.
We found  exact solutions for the case $\delta=0$, these  are $\varphi=-2\tanh\omega$, $\alpha=0$; and
\[\varphi=\pm\frac{1-12\varepsilon}{10\varepsilon}-\frac{12}5\tanh\omega\pm\frac65\tanh^2\omega,\quad \alpha=\pm\frac1{10\varepsilon}.\]
Thus, equation $u_t+ uu_x-u_{xx}+\varepsilon  u_{xxt}=0$ admits the exact solutions $u=-2\tanh x$ and
\[u=\pm\frac{1-12\varepsilon}{10\varepsilon}-\frac{12}5\tanh\left(x\pm\frac t{10\varepsilon}\right)\pm\frac65\tanh^2\left(x\pm\frac t{10\varepsilon}\right).\]

\section{Conclusion and discussion}

Using the method of mapping between classes we have presented the complete group classification of BBMB equations~\eqref{bbmb_fghk}. As a by-product of this approach we  also got the group classification of a related class of BBMB equations with a forcing term~\eqref{eqBBMimaged}. For the convenience of applications we adduced the results in two ways: the classification list where only inequivalent equations are presented (Table~2) and the list with their most general forms (Table~3).
Lie symmetries of the BBMB equations are not very reach. The maximal Lie invariance algebras of such equations are at most two-dimensional.
Nevertheless, the derived group classification  has revealed equations~\eqref{bbmb_fghk} that may be of more potential interest for applications and  for which the classical  Lie reduction method can be used. The classification of reductions was also performed as well as some exact solutions were constructed.

We note that
each equation from the class~\eqref{bbmb_fghk} admits the ``natural'' conservation law with the constant characteristic $\lambda=1$.
The corresponding density and flux are
\[
F^1=u,\quad G^1=f(t)u+\frac12g(t)u^2+k(t)u_x+h(t)u_{tx}.
\]
For general admitted values of the arbitrary elements~$f$, $g$, $k$, and~$h$
the associated space of conservation laws is one-dimensional.
The interesting problem is to classify local conservation laws of equations from class~\eqref{bbmb_fghk} and to find those values of arbitrary elements for which the dimension of space of conservation laws will be higher. Conservation laws lead to auxiliary (potential) systems which admit Lie symmetries that might induce nonlocal (potential) symmetries of the class~\eqref{bbmb_fghk}. The auxiliary system that corresponds to the above conservation law admits Lie symmetries that project into Lie symmetries of equations~\eqref{bbmb_fghk} and therefore do not induce nontrivial potential symmetries of these equations.

{\small
\subsection*{Acknowledgments}{
OV  would like to thank  the Department of Mathematics, FNSPE, Czech Technical University in Prague and the Department of Mathematics and Statistics of the University of Cyprus for the hospitality and support.}

\bibliographystyle{abbrv}
\bibliography{vcBBM}}

\end{document}